\documentclass[showpacs,aps,prd,nofootinbib,superscriptaddress]{revtex4}
\usepackage{graphicx}
\usepackage{dcolumn}
\usepackage{amsmath}
\usepackage{epsfig}
\usepackage{colordvi}
\usepackage{color}
\usepackage{hhline}

\begin{document}
\title{\large \bf \boldmath
Study of the reaction $e^+e^- \to \pi^0\gamma$ with the SND detector
at the VEPP-2M collider}

\author{M.~N.~Achasov}
\author{K.~I.~Beloborodov}
\author{A.~V.~Berdyugin}
\affiliation{Budker Institute of Nuclear Physics, SB RAS, Novosibirsk, 630090, Russia}
\affiliation{Novosibirsk State University, Novosibirsk, 630090, Russia}
\author{A.~G.~Bogdanchikov},
\affiliation{Budker Institute of Nuclear Physics, SB RAS, Novosibirsk, 630090, Russia}
\author{T.~V.~Dimova}
\author{V.~P.~Druzhinin}
\author{V.~B.~Golubev}
\affiliation{Budker Institute of Nuclear Physics, SB RAS, Novosibirsk, 630090, Russia}
\affiliation{Novosibirsk State University, Novosibirsk, 630090, Russia}
\author{I.~A.~Koop}
\affiliation{Budker Institute of Nuclear Physics, SB RAS, Novosibirsk, 630090, Russia}
\affiliation{Novosibirsk State University, Novosibirsk, 630090, Russia}
\affiliation{Novosibirsk State Technical University,Novosibirsk, 630092,Russia}
\author{A.~A.~Korol}
\affiliation{Budker Institute of Nuclear Physics, SB RAS, Novosibirsk, 630090, Russia}
\affiliation{Novosibirsk State University, Novosibirsk, 630090, Russia}
\author{S.~V.~Koshuba}
\author{E.~V.~Pakhtusova}
\affiliation{Budker Institute of Nuclear Physics, SB RAS, Novosibirsk, 630090, Russia}
\author{S.~I.~Serednyakov}
\author{Yu.~M.~Shatunov}
\author{Z.~K.~Silagadze}
\affiliation{Budker Institute of Nuclear Physics, SB RAS, Novosibirsk, 630090, Russia}
\affiliation{Novosibirsk State University, Novosibirsk, 630090, Russia}
\author{A.~N.~Skrinsky}
\affiliation{Budker Institute of Nuclear Physics, SB RAS, Novosibirsk, 630090, Russia}
\author{Yu.~V.~Usov}
\author{A.~V.~Vasiljev}
\affiliation{Budker Institute of Nuclear Physics, SB RAS, Novosibirsk, 630090, Russia}
\affiliation{Novosibirsk State University, Novosibirsk, 630090, Russia}
\collaboration{The SND Collaboration}

\begin{abstract}
The process $e^+e^- \to \pi^0\gamma$ has been studied  with the SND detector 
at the VEPP-2M $e^+e^-$ collider.
The $e^+e^- \to \pi^0\gamma$ cross section has been measured in 
the center-of-mass energy range from 0.60 to 1.38 GeV.
The cross section is well described by the vector meson 
dominance model. From the fit to the cross section data we have determined
the branching fractions
$B(\rho\to\pi^0\gamma)=(4.20\pm0.52)\times10^{-4}$,
$B(\omega\to\pi^0\gamma)=(8.88\pm0.18)\%$,
$B(\phi\to\pi^0\gamma)=(1.367\pm0.072)\times10^{-3}$,
and the relative phase between the $\rho$ and $\omega$ amplitudes
$\varphi_{\rho}=(-12.7\pm4.5)^\circ$.
Our data on the process $e^+e^- \to \pi^0\gamma$ 
are the most accurate to date.
\end{abstract}
\pacs{13.66.Bc, 14.40.Be, 13.20.Gd, 13.40.Gp}

\maketitle

\section{Introduction}
Nowadays, great attention is paid to both experimental and theoretical 
studies of the $\pi^0\gamma^{(\ast)}\gamma^{(\ast)}$ transition form factor. 
This interest is mainly due to the problem of calculating the hadronic 
light-by-light contribution to the value of the muon anomalous magnetic 
moment $(g-2)_\mu$~\cite{lbl}.
The theoretical uncertainty of this contribution is responsible for a 
sizable part of the uncertainty of the $(g-2)_\mu$ calculation. 
Experimental data on the form factors 
needed for development of phenomenological models are derived from 
measurements of two-photon $\pi^0$ production 
$e^+e^- \to e^+ e^-\gamma^\ast \gamma^\ast \to e^+ e^- \pi^0$, two-photon and 
conversion decays 
$\pi^0\to\gamma\gamma$, $\gamma e^+ e^-$, and $e^+ e^- e^+ e^-$, 
and the radiative process $e^+ e^- \to \gamma^\ast\to\pi^0 \gamma$.
Investigations of these processes are planned in various
experiments (Belle-2, BES-III, KLOE-II, SND, CMD-3).

From analysis of the $e^{+}e^{-}\rightarrow \pi ^{0}\gamma$ data in the 
vector meson dominance (VMD) model, the widths of radiative decays of vector 
mesons can be extracted. The values of these probabilities for low-lying vector 
resonances $\rho(770)$, $\omega(782)$ and $\phi(1020)$ are widely used in
phenomenological models, in particular, to fix their quark content. 
The radiative decays of the excited states of light mesons have been studied
very badly.
Since these probabilities are sensitive to the quark content of the mesons,
their measurements are important to search for exotic states
(glueballs, hybrid mesons), which are predicted in the mass range between
1 and 2 GeV.

The most accurate studies of the  process $e^{+}e^{-}\rightarrow \pi ^{0}\gamma$
were performed in experiments at the VEPP-2M $e^+e^-$ collider
with the SND~\cite{snd1,snd2} and CMD-2~\cite{cmd} detectors.
From these data, only $\omega\to \pi^0 \gamma$ decay has been measured with a 
relatively high accuracy. The combined SND and CMD-2
result on the product $B(\omega \to \pi^0 \gamma )B(\omega \to e^+e^-)$
has an uncertainty of 2.6\%. However, the value of this product 
differs by 7\% from that calculated using 
$B(\omega \to \pi^0 \gamma )$ and $B(\omega \to e^+e^-)$ given
in the Particle Data Group (PDG) table~\cite{pdg}.
This difference is caused by existing contradictions between the measured 
values of
$B(\omega \to \pi^0 \gamma )B(\omega \to e^+e^-)$,
$B(\omega \to e^+e^-)B(\omega \to \pi^+\pi^-\pi^0)$, and
$B(\omega \to \pi^0 \gamma )/B(\omega \to \pi^+\pi^-\pi^0)$.
The two latter parameters have accuracies of
1.6\% and 1.8\%, respectively, and determine the current 
PDG value of $B(\omega \to \pi^0 \gamma )$.
To resolve or enhance this contradiction, it is necessary to improve the 
accuracy of the $e^+e^- \to \pi^0 \gamma$ cross-section measurement
near the $\omega$-resonance peak. 

The accuracy of the $\rho \to \pi^0 \gamma$ branching fraction (13\%) is 
determined by statistics of existing measurements. The formal accuracy 
of the PDG value for the $\phi \to \pi^0 \gamma$ branching fraction is better 
than 5\%.
This PDG value is obtained by averaging the measurements~\cite{snd1,cmd}
with a systematic error of about 8\% each. The systematic error arises from 
the uncertainty in the nonresonant amplitude interfering with the amplitude of 
the $\phi \to \pi^0 \gamma$ decay. The nonresonant amplitude is determined 
by the tails of the $\rho$ and $\omega$ resonances, as well as by 
the contributions of higher excitations of the vector resonances. 
To reduce this 
systematic error, it is necessary to improve the accuracy of the 
$e^+e^- \to \pi^0 \gamma$
cross section in the wide energy range, from 0.6 up to at least 1.4 GeV.

It should be noted that the published SND results are based on about 25\% 
of data collected at VEPP-2M. 
In this work the full data sample recorded by SND at VEPP-2M is used
to measure the $e^+e^- \to \pi^0 \gamma$ cross section in the energy range 
from 0.6 to 1.4 GeV.

\section{Experiment} \label{Exper}
   SND~\cite{snd} is a general-purpose non-magnetic detector. 
In the period from 1996 to 2000 it collected data at the VEPP-2M $e^+e^-$
collider~\cite{vepp2m}. The main part of the detector is a spherical
three-layer calorimeter containing 1640 individual NaI(Tl) crystals. 
The calorimeter covers a solid angle of 95\% of $4\pi$; its 
thickness for particles coming from the collider interaction region is
$ 13.4\, X_{0} $. The calorimeter energy resolution for
photons is $\sigma_{E}/E_\gamma=4.2\%/\sqrt[4]{E_\gamma({\rm GeV)}}$, the angular
resolution $\simeq 1.5^\circ$. Directions of charged particles are measured 
by a system of two cylindrical drift chambers.
Outside the calorimeter a muon detector is located, which consists of
plastic scintillation counters and streamer tubes. In this analysis
the muon detector is used as a cosmic-ray veto.

The analysis presented in this paper is based on SND data with an integrated 
luminosity of 26 pb$^{-1}$ collected in 1997--2000 in the c.m. energy 
range $\sqrt{s}=0.36-1.38$ GeV. The data were recorded during several c.m.
energy scans listed in Table~\ref{tab1}.
\begin{table*}
\caption{The SND experiments used in this analysis.\label{tab1}}
\begin{ruledtabular}
\begin{tabular}{ccc}
Year & C.m. energy range (MeV) & Integrated luminosity (pb$^{-1}$) \\
1997 & 1060--1380 & 5.7 \\
1998 &  984--1060 & 7.8 \\
1998 &  360--970  & 3.5 \\
1999 & 1060--1360 & 3.0 \\
2000 &  600--940  & 5.9 \\
\end{tabular}
\end{ruledtabular}
\end{table*}
The step of the scans varied from 0.5 MeV near the peaks of the 
$\omega$ and $\phi$ resonances to 10--20 MeV far from the
``narrow`` resonances. 

The beam energy is calculated from the 
information about the magnetic field value in the bending
magnets and revolution frequency of the collider recorded during experiment.  
The relative 
accuracy of the energy setting for each energy point is about 50 keV,
while the common shift of the energy scale within 
the scan can amount to 0.5 MeV. At three energy points in the 
vicinity of the $\omega$-resonance the beam energy was
measured using the resonant depolarization method~\cite{depol}.
The accuracy of the center-of-mass energy calibration is 0.04 MeV.
These measurements allowed to fix the energy scale in the
1998 scan of the $\rho-\omega$ region. The scale for the 2000 scan
was calibrated using the $\omega$-mass measurement in the
process $e^+e^-\to\pi^+\pi^-\pi^0$~\cite{SNDom3pi}.
In the vicinity of the $\phi$ resonance
the beam energy was measured using charged kaons from $\phi\to K^+K^-$ 
decay detected by the CMD-2 detector which took data simultaneously with SND.
The accuracy of the energy scale calibration near the $\phi$ is 
estimated to be 0.04 MeV.

For simulation of signal events, we use an event generator, 
which includes radiative corrections to the Born cross section 
calculated according to Ref.~\cite{FadinRad}. In particular, an extra 
photon emitted
by initial electrons is generated with the angular distribution 
taken from Ref.~\cite{BoneMartine}. 
The event generator for  the process $e^+e^-\to \gamma\gamma$ used 
for normalization is based on Ref.~\cite{Berends}. The theoretical 
uncertainty of the $e^+e^-\to \gamma\gamma$ cross section calculation
is estimated to be 1\%. Simulation takes into account
variations of experimental conditions during data taking, in particular,
dead detector channels, and beam-generated background. Due to the beam 
background, some part of data events contains spurious tracks and 
photons. 
To take into account this effect in MC simulation, beam-background events
recorded during experiment with a special random trigger are merged with
simulated events. 

\section{Event selection\label{evsel}}
In this analysis, we simultaneously select three-photon events of
the process under study 
$e^+e^-\to\pi^0\gamma\to 3\gamma$
and two-photon events of the process
$e^+e^-\to \gamma\gamma$ used for normalization.
Some selection criteria, such as absence of
charged tracks in an event and the muon-system veto, 
are common for both processes. So, 
systematic uncertainties associated with these 
criteria cancel as a result of the normalization.

Two- and three-photon data events
must satisfy  the following first-level-trigger (FLT) conditions.
There are at least two clusters in the calorimeter
with the energy deposition larger than 30 MeV, 
no tracks found by the FLT track finder in the tracking system, 
and no signal in the muon system. 
The total energy deposition in the calorimeter should exceed a threshold,
which varies with energy, but is always below $ 0.4\sqrt{s} $.

The preliminary selection criteria for reconstructed events are also 
the same for the processes $e^+e^-\to \gamma\gamma$ and $e^+e^-\to 3\gamma$.
There are no charged particles in an event. The total 
energy deposition in the calorimeter is required to be larger than 
$0.65\sqrt{s}$, and the total event momentum calculated using energy 
deposition in the calorimeter crystals should be less than $0.3\sqrt{s}$.

\begin{figure}
\includegraphics[width=0.4\textwidth]{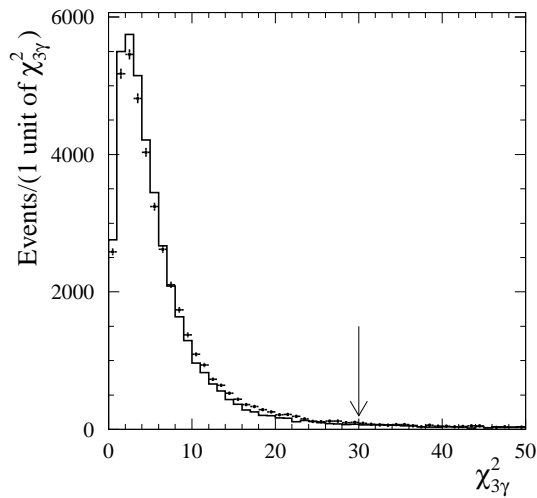}
\caption{The $\chi^2_{3\gamma}$ distribution for data (points with
error bars) and simulated $e^+e^-\to \pi^0\gamma$ events (histogram)
from the energy region near the $\omega$-resonance ($779<\sqrt{s}<787$ MeV).
The distributions are normalized to the same area.
The arrow indicates the upper limit of the selection condition.
\label{fig1}}
\end{figure}

The $e^+e^-\to \pi^0\gamma$ candidate events should have at least
three reconstructed photons with the energy larger than 50 MeV.
For these events we perform a kinematic fit with
four constraints of energy and momentum balance.
For events with more than three photons, the fit uses parameters of 
three photons with highest energy. The distribution of $\chi^2$
of the kinematic fit ($\chi^2_{3\gamma}$) is shown in Fig.~\ref{fig1}.
In the energy region below 1.06 GeV we select events with 
$\chi^2_{3\gamma}<30$. In the region $\sqrt{s}>1.06$ GeV, where 
the signal-to-background ratio is low,  a tighter condition
$\chi^2_{3\gamma}<20$ is used. For further selection we use
the fitted parameters of the three photons.
Their polar angles are required to be
in the range $ 36^{\circ }<\theta_\gamma<144^{\circ } $. In the
energy region $\sqrt{s}>1.06$ additional conditions are used.
To remove background from five-photon events of the
process $e^+e^-\to\omega\pi^0\to\pi^0\pi^0\gamma$, it is
required that the number of photons in an event be exactly
three. The condition on the photon energy $E_\gamma>0.125\sqrt{s}$ 
is applied to increase the signal-to-background ratio. 

\begin{figure}
\includegraphics[width=0.4\textwidth]{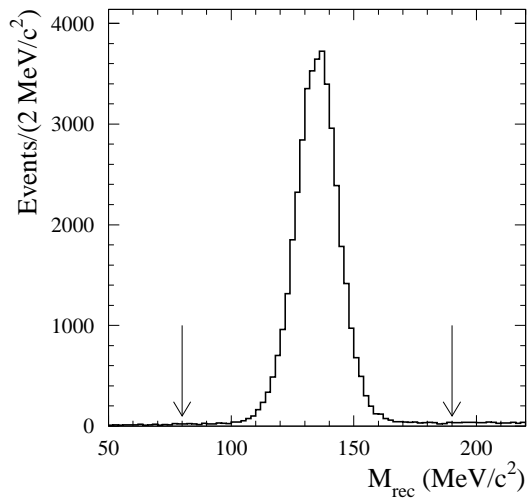}
\caption{The $M_{\rm rec}$ distribution for data events 
with $\chi^2_{3\gamma}<30$ from the energy region near 
$\omega$-resonance ($779<\sqrt{s}<787$ MeV).
The arrows indicate the selection criterion 
$80<M_{\rm rec}<190$ MeV/$c^2$. \label{fig2}}
\end{figure}

The distribution of the mass recoiling against the most energetic 
photon in an event ($M_{\rm rec}$) is shown in Fig.~\ref{fig2}.
We select events with $80<M_{\rm rec}<190$ MeV/$c^2$.

Two-photon events of the process $e^+e^-\to \gamma\gamma$ are selected 
with the following selection criteria. There are at least two photons in
an event with $E_\gamma>0.3\sqrt{s}$. The azimuthal and polar angles of 
these photons satisfy the conditions 
$| |\phi_1-\phi_2| - 180^{\circ}| < 15^{\circ}$,
$|\theta_1+\theta_2 - 180^{\circ}| < 20^{\circ}$,
and
$180^{\circ}-|\theta_1-\theta_2| >45^{\circ}  $.

\section{Fitting the $M_{\rm rec}$ spectra\label{mrecfit}}
To determine the number of signal events ($N_{\rm sig}$), 
the $M_{\rm rec}$ spectrum
is fitted by a sum of signal and background distributions.
The signal distribution is described by a double-Gaussian function.

The sources of background for the process under study are 
$e^+e^-\to 3\gamma$ events, and $e^+e^-\to \gamma\gamma$ events with 
a fake photon arising from the beam background or splitting of electromagnetic 
showers. In the energy region near the $\phi(1020)$ resonance 
the process $e^+e^-\to \eta\gamma$ should be 
also taken into account. It increases background by about 40\% in
the resonance peak. 
The background composition outside the $\phi$-meson
region is 30\% from $e^+e^-\to \gamma\gamma$ 
and 70\% from $e^+e^-\to 3\gamma$. 
All background processes have the $M_{\rm rec}$ distribution not peaked 
near the $\pi^0$ mass.
The simulation shows
that the shape of the $M_{\rm rec}$ distribution in the chosen
mass window $80<M_{\rm rec}<190$ MeV/$c^2$ for
the processes $e^+e^-\to \gamma\gamma$ and $e^+e^-\to 3\gamma$ 
is close to linear in the range $0.6 \le \sqrt{s} \le 1.06$ GeV. 
Above 1.06 GeV it is well described by a second-order polynomial. 
In the energy range $0.36 \le \sqrt{s}\le 0.58$ GeV, 
the $M_{\rm rec}$ distribution for $e^+e^- \to 2(3)\gamma$ background 
events has a maximum in the chosen 
$M_{\rm rec}$ window.
An inaccuracy of background-shape simulation in this energy range
may be a serious source of the systematic uncertainty in determination of the 
number of signal events. We don't see any clear $\pi^0$ signal over 
background in the $M_{\rm rec}$ spectrum for energies below 0.6 GeV. Since the
total integrated luminosity collected at eight energy points between 
0.36 and 0.58 GeV is lower than that for the energy point $\sqrt{s}=0.6$ GeV, 
we exclude these points from the current analysis.

The simulation predicts the number of background events with an accuracy 
better than 5\% in the $\phi(1020)$
region and below, and with an accuracy of about 10--15\% above.
In the fit to the $M_{\rm rec}$ spectrum, the background distribution 
is described by the distribution obtained from simulation plus a linear 
function. The latter is needed to take into account a difference between 
data and simulation in the shape of the background distribution and in 
the number of background events.

At the energy points with large $\pi^0$ statistics ($N_{\rm sig}>3000$)
the fit is performed with 8 free parameters (6 for signal and 2 
for background). At the points with lower statistics, where the fit with
floating double-Gaussian parameters is unstable, the signal distribution is 
obtained by fitting the mass spectrum for simulated signal events.
To take into account a difference between data and MC simulation in mass
calibration and resolution, the signal distribution obtained
in simulation is modified in the following way:
$M_{\rm rec}^{\rm data}=M_{\rm rec}^{\rm MC}+\Delta M$, and 
$\sigma^2_{\rm data}=\sigma^2_{\rm MC}+\Delta_{\sigma^2}$ for
both $\sigma$'s of the double-Gaussian function. The parameters 
$\Delta M$ and $\Delta_{\sigma^2}$ are determined by fitting the 
data and simulated $M_{\rm rec}$ spectra in the
energy region near the $\omega$ resonance. They are found to
be $\Delta M=-0.63\pm0.05$ ($-0.84\pm0.03$) MeV/$c^2$ and
$\Delta_{\sigma^2}=0.5\pm0.7$ ($-1.2\pm0.5$) MeV$c^2$/$c^4$
for the 1998 (2000) energy scan. The values obtained for the 1998 
scan of the $\omega$ region are used for analysis of data collected in 
1997-1998, while the 2000 values for analysis of 1999-2000 scans.
A possible systematic uncertainty due to using the simulated $M_{\rm rec}$ 
spectrum is estimated by comparing the two fitting methods at the energy points
with $N_{\rm sig}>3000$. It is found to be less than 0.2\% and is negligible
compared with the statistical error of $N_{\rm sig}$.

\begin{figure*}
\includegraphics[width=0.32\textwidth]{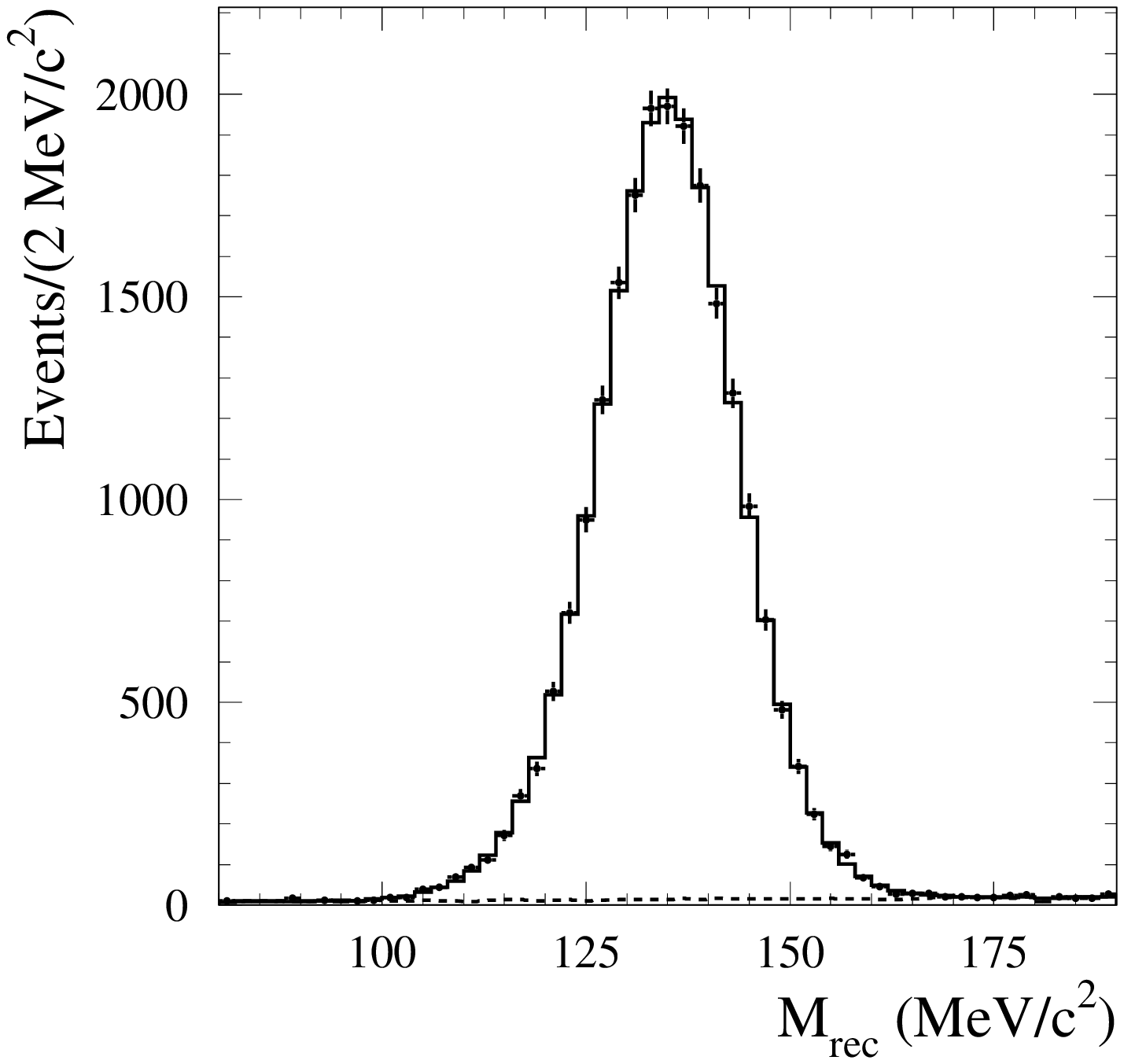}
\includegraphics[width=0.32\textwidth]{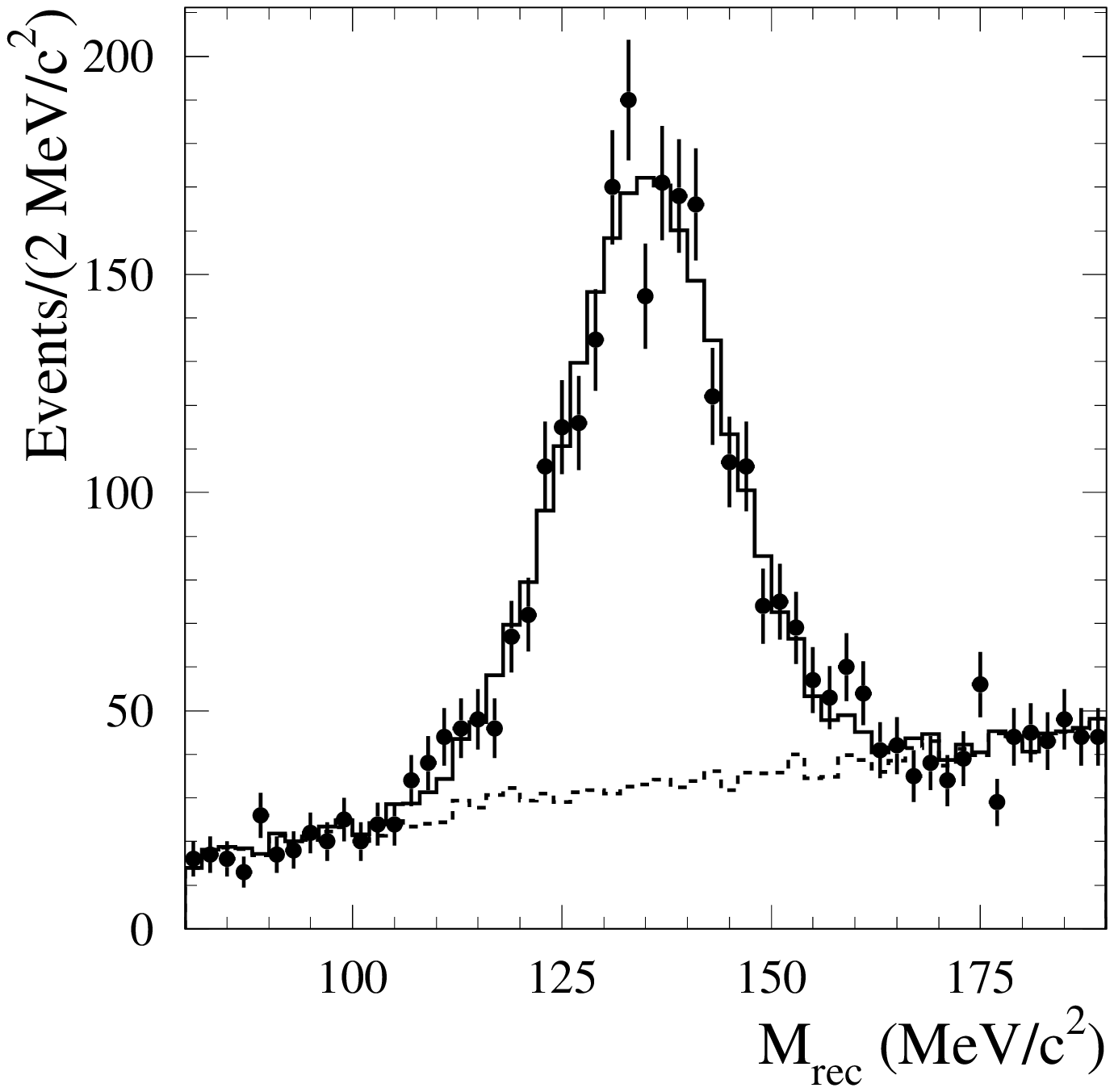}
\includegraphics[width=0.32\textwidth]{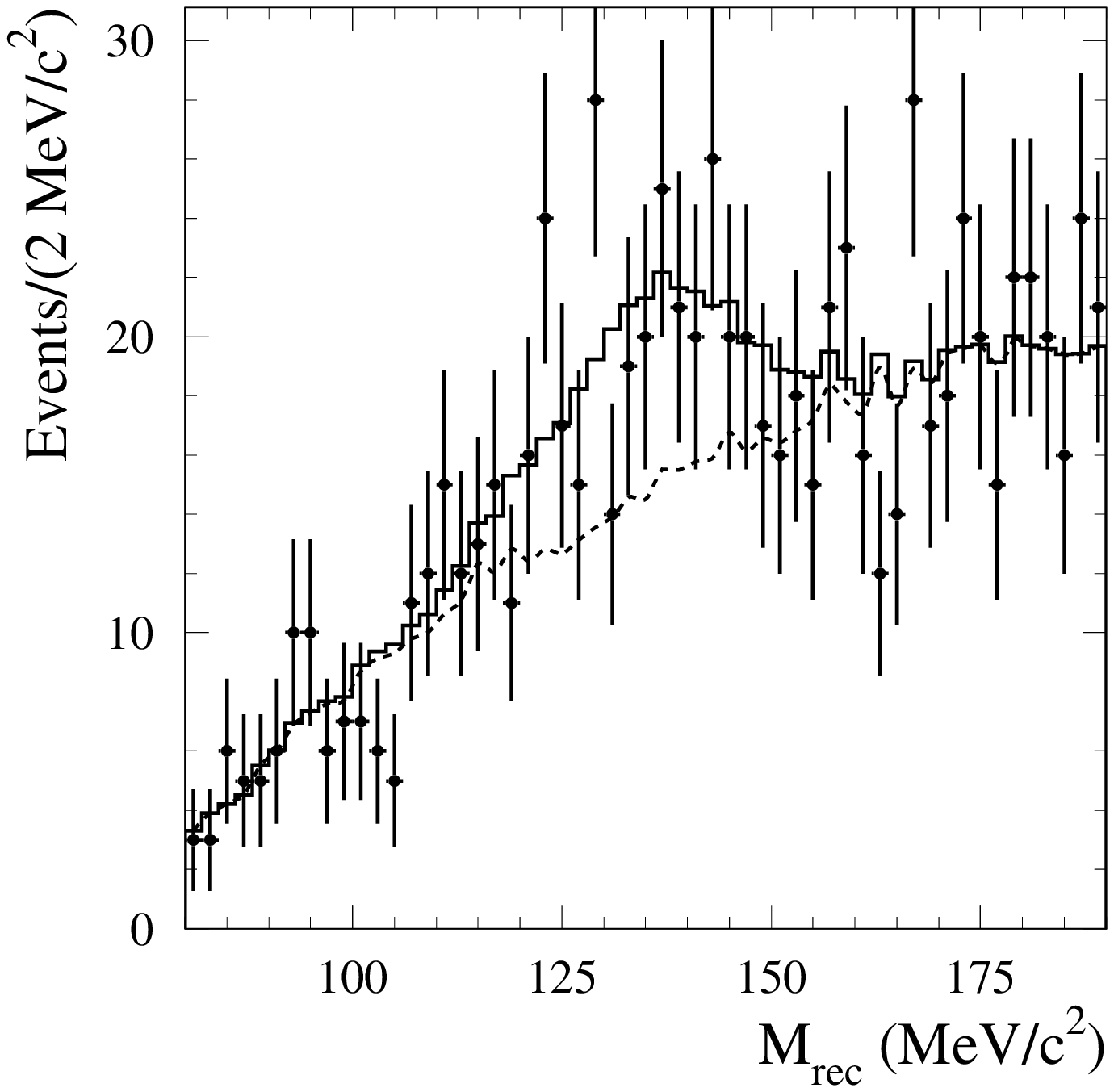}
\caption{The $M_{\rm rec}$ distribution for data events (points with error 
bars) with $\sqrt{s}=783.35$ GeV (left), $\sqrt{s}=1018.7$ GeV (middle), and
$\sqrt{s}>1.06$ GeV (right). The solid histogram represents the result of
the fit described in the text. The dashed histogram shows the fitted
background distribution.
\label{fig3}}
\end{figure*}

  The results of the fit at the energy points near
the peaks of the $\omega$ and $\phi$ resonances, and in the region
$\sqrt{s} > 1.06$ GeV are shown in Fig.~\ref{fig3}. 
The obtained numbers of signal events for different energy points
are listed in Table~\ref{allres}. Since the cross section values obtained
for the 1998 and 2000 energy scans are found to be statistically compatible,
data samples for the two scans in energy points located far from the $\omega$ 
region ($600 \le \sqrt{s}<765$ MeV and $800<\sqrt{s}<945$ MeV) are combined.  

The fitted number of signal events in the energy region $1.06 <\sqrt{s} <1.38$
GeV is $97\pm24$. This energy region
is separated into five subintervals. Data of the 
1997 and 1999 scans are combined. The boundaries of the subintervals,
the average subinterval energies, calculated as $\sum \sqrt{s_i}L_i/\sum L_i$,
where $\sqrt{s_i}$ and $L_i$ are the energy and integrated luminosity for the
$i$th energy point included in the subinterval,
and the fitted numbers of signal events are listed in the last five rows of 
Table~\ref{allres}. 

\section{Luminosity measurement\label{lum}}
In the energy region of the $\omega$ resonance, two-photon events selected 
using the criteria described in Sec.~\ref{evsel} contain a significant
fraction of $e^+e^-\to \pi^0\gamma$ events (up to 20\% at the resonance
peak). To subtract this resonance background, the two-photon events 
are divided into two classes:
satisfying (class I) and not satisfying (class II) the conditions
$\chi^2_{3\gamma}<100$ and $80<M_{\rm rec}<190$ MeV/$c^2$.
For class-I events, which contain three reconstructed photons, we fit to the
$M_{\rm rec}$ spectrum using the fitting function described in 
Sec.~\ref{mrecfit}, and determine the number of events ($N_{2\gamma,1}$) not
peaked at the $\pi^0$ mass. The number of two-photon events at each 
energy point used for a luminosity measurement is calculated as follows
\begin{equation}
N_{2\gamma}=N_{2\gamma,1}+N_{2\gamma,2}
-N_{\rm sig}\frac{\varepsilon_{\pi^0\gamma}^{2\gamma}}
{\varepsilon_{\pi^0\gamma}^{\pi^0\gamma}},
\label{nlum}
\end{equation}
where the second and third terms are the number of events and
the estimated $e^+e^-\to \pi^0\gamma$ background in class II, respectively.
In the third term, $N_{sig}$ is the number of $e^+e^-\to \pi^0\gamma$ events 
determined
in Sec.~\ref{mrecfit}, and $\varepsilon_{\pi^0\gamma}^{\pi^0\gamma}$ and 
$\varepsilon_{\pi^0\gamma}^{2\gamma}$ are the detection efficiencies 
determined using $e^+e^-\to \pi^0\gamma$ simulation for the $\pi^0\gamma$ and
$2\gamma$ (class II) selection criteria. The third term is about 8\% of 
$N_{2\gamma}$ in the maximum of the $\omega$ resonance.

The quality of background subtraction is tested by analyzing the
energy dependence of the $N_{2\gamma}/N_{e^+e^-}$ ratio, where 
$N_{e^+e^-}$ is the number of selected $e^+e^-\to e^+e^-$ events. 
Selection of $e^+e^-\to e^+e^-$ events described
in detail in Ref.~\cite{sndpipi} fully removes background from
$\omega$ decays. The $N_{2\gamma}/N_{e^+e^-}$ energy dependence is
fitted by a sum of a linear function and a Breit-Wigner function 
describing the $\omega$-resonance contribution. The resonance background
fraction is found to be $(0.3\pm0.3)\%$ at the $\omega$ peak.
To take into account the contribution of this resonance, we multiply the  
$\varepsilon_{\pi^0\gamma}^{2\gamma}/\varepsilon_{\pi^0\gamma}^{\pi^0\gamma}$
ratio used in Eq.~(\ref{nlum}) by a factor of $1.04\pm0.04$. 

A similar procedure is used to subtract $e^+e^-\to \pi^0\gamma$ background 
in the $\phi$-meson energy region. It is found to be less than that in the
$\omega$ energy region by a factor of 20. Another source of background near
the $\phi$-resonance is the decay chain $\phi\to \eta\gamma\to 3\gamma$.
To suppress the $\eta\gamma$ background by a factor of about 4, the additional
cut $E_{\gamma,min} < 0.125\sqrt{s}$ is applied in the energy region
$0.984<\sqrt{s}<1.060$ GeV for events with $\chi^2_{3\gamma}<100$, where  
$E_{\gamma,min}$ is the energy of the third, less energetic photon in an event.
With this cut the fraction of the $\eta\gamma$ background does not exceed 
0.5\%. Total resonance background in the $\phi$-meson energy region is about
0.8\%. We estimate that the uncertainty associated with subtraction of this 
background is negligible.

The integrated luminosity calculated as $L=N_{2\gamma}/\sigma_{2\gamma}$ is
listed in Table~\ref{allres}, where $\sigma_{2\gamma}$ is the 
$e^+e^-\to\gamma\gamma$ cross section calculated using MC simulation for
the selection criteria described in Sec.~\ref{evsel}. For most energy
points the statistical error of the integrated luminosity does not exceed 1\%.
The systematic uncertainty is 1.2\% and includes the theoretical error of
cross section calculation (1\%) and uncertainty associated with a difference
between data and simulation in photon angular and energy resolutions (0.7\%).
The latter is estimated by variation of the boundaries of the
angular and energy cuts used for selection of two-photon events.
The main contribution to this uncertainty comes from the condition
on the photon polar angles $180^{\circ}-|\theta_1-\theta_2| > 45^{\circ}$.
The uncertainties associated with the conditions common for the
two-gamma and three-gamma selections (cosmic-ray veto, absence of charged 
tracks, etc.) cancel in the $N_{\rm sig}/N_{2\gamma}$ ratio and are not 
included in the  systematic error quoted above.

\section{Detection efficiency and radiative corrections}
The visible cross section for the process $e^+e^- \to \pi^0\gamma$
is written as
\begin{equation}
\sigma_{vis}(s) = \int \limits_{0}^{x_{max}}\varepsilon_r(s, x)F(x,s)
\sigma(s(1-x))dx,
\label{viscrs}
\end{equation}
where $\sigma(s)$ is the Born cross section extracted
from the experiment, $F(x,E)$ is a so-called radiator function describing the
probability to emit from the initial state extra photons with the total energy
$x\sqrt{s}/2$~\cite{FadinRad}, $x_{max}=1-m_{\pi^0}^2/s$, and 
$\varepsilon_r(s, x)$ is the detection efficiency. The detection
efficiency is determined using MC simulation, as a function of $\sqrt{s}$ 
and $x=2E_r/\sqrt{s}$, where $E_r$ is the energy of the extra photon emitted
from the initial state. It is parametrized as 
$\varepsilon_r(s, x)=\varepsilon(s)g(s,x)$, where 
$\varepsilon(s)\equiv \varepsilon_r(E,0)$.
We use the approximation when all variations of experimental conditions 
(dead calorimeter channels, beam background, etc.) are accounted for in 
$\varepsilon(s)$, while $g(s,x)$ is a smooth function of $\sqrt{s}$. With this
parametrization, Eq.~(\ref{viscrs}) can be rewritten in the conventional form:
\begin{equation}
\label{viscrs2} \sigma_{vis} = \varepsilon(s)\sigma(s)(1+\delta(s)),
\end{equation}
where $\delta(s)$ is the radiative correction.

\begin{figure}
\includegraphics[width=0.4\textwidth]{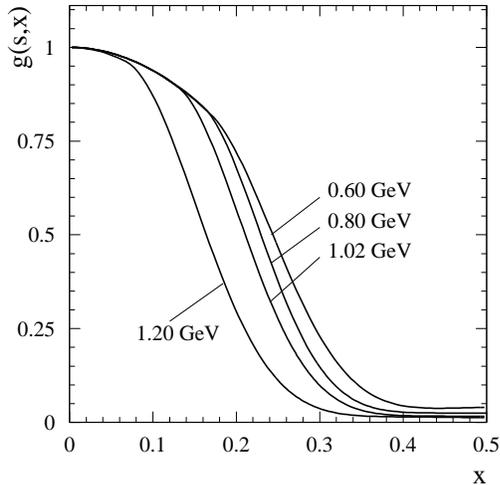}
\caption{The $x$ dependence of the detection efficiency obtained
from simulation for four $\sqrt{s}$ values.
\label{fig4}}
\end{figure}

The functions $\varepsilon(s)$ and $g(s,x)$ are determined using MC
simulation. Since the standard $e^+e^-\to \pi^0\gamma$ event generator
has the $dN/dx$ distribution proportional to $1/x$, a special sample 
of simulated $\pi^0\gamma$ events with $dN/dx = {\rm const}$ has been
produced to increase statistics at large $x$.
The obtained $x$ dependence of the detection efficiency is approximated 
by a smooth function. The result of 
the approximation for four representative $\sqrt{s}$ values is shown in 
Fig.~\ref{fig4}.
Dependence of the $g(s,x)$ shape on $s$ is not strong. In the energy range 
$0.60<\sqrt{s} < 1.06$ GeV, where the same cut $\chi^2_{3\gamma}<30$ is used,
the effective threshold ($x_{th}$) determined from the equation
$g(s,x_{th})=0.5$ changes from 0.21 to 0.24. At higher energies, where
we use a tighter cut $\chi^2_{3\gamma}<20$, $x_{th}$ is about 0.16. 

The detection efficiency $\varepsilon_{\rm MC}(s)$ determined using MC 
simulation is corrected to take into account a difference between data and 
simulation in the detector response 
\begin{equation}
\varepsilon(s)=\varepsilon_{\rm MC}(s)\prod(1+\delta_i),
\end{equation}
where $\delta_i$ is the efficiency correction discussed below.
To determine the efficiency corrections and estimate systematic uncertainties 
due to imperfect simulation of the detector response for photons, 
we study data and simulated signal events in the narrow energy range near 
the peak of the $\omega$ resonance, $0.777<\sqrt{s}<0.785$ GeV, where the 
signal-to-background ratio in the mass window $80<M_{\rm rec}<190$ MeV/$c^2$ is 
about 25 for our standard selection criteria.

To estimate the systematic uncertainty associated with the condition on photon 
polar angles ($\theta_0<\theta_\gamma<180^{\circ }-\theta_0 $ with 
$\theta_0=36^{\circ }$), we vary $\theta_0$ from $27^{\circ }$ to
$45^{\circ }$. The range of the $\theta_0$ variation corresponds to a doubled 
angular size of the calorimeter crystal. The number of selected events
increases (decreases) by 22 (24)\% for $\theta_0=27^{\circ }$ ($45^{\circ }$),
while the maximum deviation of the visible cross section obtained at
different $\theta_0$ from that for $\theta_0=36^{\circ }$ does not 
exceed 0.6\%. This deviation is taken as an estimate of a systematic
uncertainty due to the condition on photon polar angles.

With the conditions on the total energy deposition in the calorimeter and
the total event momentum described in Sec.~\ref{evsel} all signal events have 
$\chi^2$ of the kinematic fit less than 1000. The fraction of signal events 
with $30<\chi^2_{3\gamma}<1000$ is about 5\% in the $\omega$ energy region 
defined above.
The difference between the cross sections measured in the $\omega$ energy 
region with the conditions $\chi^2_{3\gamma}<30$
and $\chi^2_{3\gamma}<1000$ is $\delta_{\chi^2}=-(0.2\pm0.2)\%$ for the 1998 scan and 
$\delta_{\chi^2}=-(1.5\pm0.2)\%$ for 
the 2000 scan. These values are used to correct the detection efficiencies 
for the  1998 and 
2000 energy scans. It should be noted that the $\chi^2_{3\gamma}$ distribution becomes 
wider with increasing energy. The condition $\chi^2_{3\gamma}<30$ near the $\phi$-meson 
resonance corresponds to $\chi^2_{3\gamma}<28$ near the $\omega$-resonance. This effect,
however, does not lead to any significant change of the correction in the 
energy region below 1.06 GeV. In the energy region $1.06-1.38$ GeV the 
condition $\chi^2_{3\gamma}<20$ is applied. The fraction of signal events with 
$20<\chi^2_{3\gamma}<1000$ varies from 7.1 to 8.0\%. This fraction corresponds to the
cut $\chi^2_{3\gamma}<(18-20)$ in the $\omega$ region. The efficiency correction is 
found to be $-(0.8\pm0.2)\%$ for the 1998 scan and $-(2.2\pm0.2)\%$ for
the 2000 scan. These corrections are used for the 1997 and 1999 scans, 
respectively, with the systematic uncertainties equal to the correction value.

In SND a photon converted in material before the tracking system is 
reconstructed as
a charged particle. Events with converted photons are rejected by 
our selection criteria. Since  the numbers of photons in the final state are 
different for the signal and normalization processes,
the data-MC simulation difference in the conversion probability leads to
a systematic shift in the measured cross section. This difference was studied 
in Ref.~\cite{phconv}, where the ratio of the conversion probabilities in 
data and simulation was found to be $0.82\pm0.04$. The loss of simulated 
events with our angular conditions due to photon conversion in 
material is 2.5\% for $e^+e^-\to \pi^0\gamma$ and 1.7\% for 
$e^+e^-\to\gamma\gamma$. The efficiency correction is calculated to be
$\delta_{\rm conv}=(0.14\pm0.03)\%$.

As it was discussed earlier, some part of the data events contains
beam-generated spurious charged tracks and photons. The effect of extra
charged tracks cancels due to normalization to two-photon events. To simulate
the beam-generated particles, events recorded during experiment with a 
special random trigger are used. These background events are 
superimposed on simulated events of the process under study. Using
$\pi^0\gamma$ events from the $\omega$ energy region, we find that the
fraction of events with extra photon(s) in data varies from 5 to 7\%. The 
difference between data and simulation in this fraction does not exceed 10\%. 
Below 1.05 GeV, where our selection criteria are weakly sensitive to the 
presence of spurious photons, there is no need in any additional 
systematic uncertainty. Above 1.05 GeV, where events with exactly three
photons are selected, a 0.7\% systematic uncertainty is additionally 
introduced. 

The total efficiency correction for $\sqrt{s}<1.06$ GeV is $(-0.1\pm 0.6)\%$
for the 1998 scan and $(-1.4\pm 0.6)\%$ for the 2000 scan. For 
$\sqrt{s}>1.06$ GeV the total correction is $(-0.7\pm 1.2)\%$ for the 1997 
scan and $(-2.1\pm 2.4)\%$ for the 1999 scan. The quoted error is the total 
systematic uncertainty of the detection efficiency. The corrected values of 
$\varepsilon$ at different 
energy points are listed in Table~\ref{allres}. The statistical error on the 
detection efficiency is negligible. A nonmonotonic behavior of 
$\varepsilon(s)$ as a function of the c.m.~energy is due to variations of 
experimental conditions. In particular, 
a fraction of dead calorimeter channels varies during experiments from
0.7\% to 1.8\%. The detection efficiency grows from 36\% at 0.6 GeV to 
43\% in the $\phi$-meson region. Above the $\phi$, where additional
selection conditions are used, it decreases from 26\% to 22\% with
increase of energy.

\begin{table}
\caption{The c.m.~energy ($E$), integrated luminosity ($L$), detection
efficiency ($\varepsilon$), number of selected signal events ($N_{\rm sig}$),
radiative-correction factor ($1+\delta$), measured Born cross section
($\sigma$). For the cross section the first error is statistical, the
second is systematic.\label{allres}}
\begin{ruledtabular}
\begin{tabular}{cccccc}
 $E$, GeV & $L$, nb$^{-1}$ & $\varepsilon$, \% & $N_{\rm sig}$ & $1+\delta$ &
 $\sigma$, nb \\[0.3ex] \hline \\[-2.1ex]
  600.00&   87& 35.3& $    0\pm 11$ &   0.919(1)& $   0   \pm0.40\pm0.01$ \\
  630.00&  118& 36.6& $   24\pm 13$ &   0.913(1)& $   0.61\pm0.33\pm0.01$ \\
  660.00&  274& 37.4& $   65\pm 19$ &   0.906(1)& $   0.70\pm0.20\pm0.01$ \\
  690.00&  172& 37.7& $   78\pm 16$ &   0.899(1)& $   1.33\pm0.28\pm0.02$ \\
  720.00&  570& 38.9& $  400\pm 32$ &   0.890(1)& $   2.03\pm0.16\pm0.03$ \\
  750.26&  221& 39.5& $  337\pm 24$ &   0.865(1)& $   4.45\pm0.32\pm0.06$ \\
  760.29&  242& 39.8& $  635\pm 30$ &   0.844(1)& $   7.81\pm0.38\pm0.11$ \\
  764.31&  253& 40.0& $  887\pm 35$ &   0.832(1)& $  10.52\pm0.42\pm0.15$ \\
  769.79&   45& 40.1& $  260\pm 18$ &   0.812(1)& $  17.88\pm1.25\pm0.28$ \\
  770.40&  243& 40.1& $ 1660\pm 45$ &   0.809(1)& $  21.02\pm0.58\pm0.31$ \\
  773.79&   64& 40.2& $  667\pm 28$ &   0.794(1)& $  32.5\pm1.4\pm0.5$ \\
  774.40&  155& 40.3& $ 1926\pm 47$ &   0.791(1)& $  39.1\pm1.0\pm0.6$ \\
  777.86&   98& 40.2& $ 2461\pm 51$ &   0.775(1)& $  80.5\pm1.8\pm1.4$ \\
  778.40&  152& 40.3& $ 4210\pm 67$ &   0.774(1)& $  89.0\pm1.5\pm1.6$ \\
  779.79&   42& 40.7& $ 1512\pm 40$ &   0.771(1)& $ 113.9\pm3.3\pm2.0$ \\
  780.20&  270& 40.3& $11063\pm108$ &   0.772(1)& $ 132.0\pm1.5\pm2.2$ \\
  780.79&  134& 40.5& $ 6202\pm 81$ &   0.774(1)& $ 147.4\pm2.2\pm2.3$ \\
  781.40&  208& 40.2& $10378\pm105$ &   0.778(1)& $ 159.8\pm1.9\pm2.4$ \\
  781.80&  377& 40.4& $19831\pm144$ &   0.782(1)& $ 166.9\pm1.5\pm2.4$ \\
  782.40&  287& 40.2& $15381\pm127$ &   0.790(1)& $ 169.2\pm1.7\pm2.3$ \\
  782.79&   83& 40.8& $ 4473\pm 69$ &   0.797(1)& $ 166.4\pm2.9\pm2.3$ \\
  783.25&  397& 40.2& $21053\pm149$ &   0.806(1)& $ 163.9\pm1.5\pm2.2$ \\
  783.79&   77& 40.8& $ 3980\pm 64$ &   0.819(1)& $ 155.8\pm2.9\pm2.2$ \\
  784.40&  276& 40.4& $13447\pm119$ &   0.836(1)& $ 144.4\pm1.5\pm2.2$ \\
  785.40&  217& 40.3& $ 9003\pm 98$ &   0.869(1)& $ 118.2\pm1.5\pm1.9$ \\
  785.87&   95& 40.6& $ 3599\pm 62$ &   0.886(1)& $ 105.5\pm2.0\pm1.7$ \\
  786.40&  172& 40.4& $ 5982\pm 88$ &   0.906(1)& $  94.8\pm1.5\pm1.6$ \\
  789.79&   58& 40.8& $ 1131\pm 35$ &   1.040(1)& $  46.1\pm1.6\pm0.7$ \\
  790.40&  133& 40.4& $ 2311\pm 51$ &   1.064(1)& $  40.4\pm1.0\pm0.6$ \\
  793.79&   54& 40.9& $  580\pm 26$ &   1.197(1)& $  22.0\pm1.2\pm0.3$ \\
  794.40&  155& 40.6& $ 1600\pm 43$ &   1.220(1)& $  20.8\pm0.7\pm0.3$ \\
  800.28&  280& 40.6& $ 1719\pm 46$ &   1.422(1)& $  10.6\pm0.4\pm0.1$ \\
  810.25&  284& 40.9& $  929\pm 36$ &   1.682(2)& $   4.76\pm0.32\pm0.07$ \\
  820.00&  320& 41.2& $  739\pm 34$ &   1.848(3)& $   3.03\pm0.26\pm0.04$ \\
  840.00&  687& 40.9& $  851\pm 42$ &   2.016(3)& $   1.50\pm0.15\pm0.02$ \\
  880.00&  383& 41.4& $  239\pm 26$ &   1.861(3)& $   0.81\pm0.16\pm0.01$ \\
  920.00&  489& 41.5& $  142\pm 26$ &   1.353(1)& $   0.52\pm0.13\pm0.01$ \\
  940.00&  480& 42.3& $   93\pm 23$ &   1.183(1)& $   0.39\pm0.11\pm0.01$ \\
  950.00&  268& 42.4& $   43\pm 15$ &   1.125(1)& $   0.33\pm0.14\pm0.01$ \\
  958.00&  241& 43.1& $   48\pm 15$ &   1.088(1)& $   0.42\pm0.14\pm0.01$ \\
  970.00&  258& 43.8& $   45\pm 15$ &   1.044(1)& $   0.38\pm0.14\pm0.01$ \\
  984.11&  353& 43.0& $   52\pm 16$ &   1.002(1)& $   0.34\pm0.11\pm0.01$ \\
 1003.82&  372& 43.0& $   67\pm 18$ &   0.905(3)& $   0.46\pm0.12\pm0.01$ \\
 1010.26&  301& 43.0& $   73\pm 16$ &   0.844(2)& $   0.67\pm0.15\pm0.01$ \\
 1015.58&  347& 43.0& $  241\pm 23$ &   0.769(1)& $   2.08\pm0.21\pm0.04$ \\
 1016.73&  595& 43.0& $  722\pm 36$ &   0.752(1)& $   3.72\pm0.19\pm0.07$ \\
 1017.66&  942& 43.0& $ 1338\pm 49$ &   0.743(1)& $   4.44\pm0.16\pm0.09$ \\
 1018.70&  986& 43.0& $ 1747\pm 55$ &   0.749(2)& $   5.63\pm0.18\pm0.08$ \\
 1019.66& 1001& 43.0& $ 1642\pm 54$ &   0.791(3)& $   4.93\pm0.16\pm0.10$ \\
 1020.53&  638& 43.0& $  893\pm 40$ &   0.871(6)& $   3.71\pm0.17\pm0.09$ \\
 1021.54&  328& 43.0& $  223\pm 23$ &   1.02(2) & $   1.53\pm0.16\pm0.05$ \\
 1022.82&  362& 43.0& $  148\pm 21$ &   1.30(5) & $   0.71\pm0.14\pm0.03$ \\
 1027.81&  369& 43.0& $   51\pm 18$ & 3.8--6.3  & $   0.05\pm0.11\pm0.03$ \\
 1033.70&  327& 43.0& $   15\pm 15$ & 8--150000 & $   0.00\pm0.11\pm0.01$ \\
 1039.62&  328& 43.0& $   10\pm 15$ & 5--15     & $   0.01\pm0.11\pm0.01$ \\
 1049.71&  365& 43.0& $   22\pm 16$ & 2.8--4.2  & $   0.04\pm0.10\pm0.01$ \\
 1059.58&  373& 43.0& $   17\pm 17$ & 2.1--2.7  & $   0.04\pm0.11\pm0.01$ \\
 1080 (1070--1090)&  780& 25.3& $   24\pm 10$ &   1.64(2) & $   0.075\pm0.049\pm0.008$ \\
 1127 (1100--1160)& 1654& 25.2& $   15\pm 12$ &   1.17(6) & $   0.030\pm0.028\pm0.002$ \\
 1201 (1180--1230)& 1659& 24.5& $   17\pm 11$ &   1.02(3) & $   0.040\pm0.027\pm0.001$ \\
 1269 (1240--1300)& 1762& 23.7& $   27\pm 10$ &   0.98(3) & $   0.065\pm0.025\pm0.003$ \\
 1350 (1310--1380)& 2781& 22.3& $   15\pm 11$ &   0.91(8) & $   0.027\pm0.019\pm0.002$ \\
\end{tabular}
\end{ruledtabular}
\end{table}

\section{Fit to cross section data}
To determine radiative corrections and the branching fractions for
the $\rho,\,\omega,\,\phi\to\pi^0\gamma$ decays, the energy dependence
of the measured visible cross section $\sigma_{vis,i}=N_{{\rm sig},i}/L_i$ is
fitted with Eq.~(\ref{viscrs}). The Born cross section is 
parametrized in the framework of the VMD model
as follows (see, for example, Ref.~\cite{FL})
\begin{eqnarray}
\sigma(s) & = & \frac{q(s)^{3}}{s^{3/2}}
\left| \sum_{V} A_V(s) \right| ^{2},\label{vsum}\\
A_{V}(s) & = & \frac{m_{V}\Gamma _{V}e^{i\varphi_V}}
{m_V^2-s-i\sqrt{s}\Gamma _{V}(s)}
\sqrt{\frac{m_{V}^{3}}{q(m_{V}^{2})^{3}}\sigma_V},\\
q(s) & = & \frac{\sqrt{s}}{2}\left( 1-\frac{m^{2}_{\pi ^{0}}}{s}\right),
\end{eqnarray}
where $ m_{V} $ is the $ V $ resonance mass, $ \Gamma _{V}(s) $
is its energy-dependent width, $\Gamma _{V}\equiv\Gamma _{V}(m_{V}^2)$,
$\varphi_V$ is the interference phase, $\sigma_V$ is the cross section
at the resonance peak, which is related to the product of the 
branching fractions for the decays $V\to e^+e^-$ and $V\to\pi^0\gamma$: 
\begin{equation}
\sigma_V=\frac{12\pi}{m_V^2}B(V\to e^+e^-)B(V\to\pi^0\gamma).
\end{equation}
The sum in Eq.(\ref{vsum}) goes over the resonances $\rho(770)$, $\omega(782)$,
$\phi(1020)$, and higher vector excitations of the $\rho$ and $\omega$ families.
The isovector and isoscalar contributions into the $e^+e^-\to \pi^0\gamma$ 
above 1.06 GeV may be estimated from the  $e^+e^-\to \omega\pi^0$ and 
$e^+e^-\to \rho\pi$ cross sections using the VMD model. In the energy 
region 1.06-1.40 GeV both contributions are found to be several tens of pb,
in reasonable agreement with the experimentally observed $e^+e^-\to \pi^0\gamma$
cross section.   
It is impossible to separate contributions to the $e^+e^-\to \pi^0\gamma$
from the $\omega(1420)$ and $\rho(1450)$ resonances, and from 
the $\omega(1650)$ and $\rho(1700)$ resonances. Therefore, in the fit we
use two effective resonances (below we will name them $V^\prime$ and
$V^{\prime\prime}$) with masses and widths of (1450,400) MeV and
(1700,300) MeV. The uncertainties of these parameters are 
assumed to be (50,50) MeV.  

The energy-dependent widths of the $\rho$, $\omega$ and $\phi$ resonances 
take into account decays with branching fractions larger than 1\%. For
the $V^\prime$ width we study two options: $\rho\pi$ phase space (dominant
for the $\omega(1420)$), and the phase space for the $\rho(1450)$, which is
a sum of the $a_1\pi$(56\%), $\omega\pi^0$(37\%), $\eta\rho$(3\%), 
and $\pi^+\pi^-$(4\%) contributions~\cite{a1pi,etapipi}. 
In the nominal fit the energy dependence of the $V^\prime$ width
is described by a half-sum of the dependences of the $\omega(1420)$ and 
$\rho(1450)$ widths. The fits with the $\omega(1420)$ and 
$\rho(1450)$ dependences are used to study a model uncertainty.
For the $V^{\prime\prime}$ width, the $\rho\pi$ phase space is used.

The phase $\varphi_\omega$ is chosen to be zero. The phases 
$\varphi_{V^\prime}$ and $\varphi_{V^{\prime\prime}}$ are set to $180^\circ$ and
$0^\circ$, respectively. Such a choice of phases for excited $\rho$ and $\omega$
states are used to describe the energy dependences of the 
$e^+e^-\to \omega\pi^0$ and $e^+e^-\to \rho\pi$ cross sections 
(see, for example, Refs.~\cite{SNDompi,SNDrhopi}).

The free fit parameters are $\sigma_\rho$, $\sigma_\omega$,
$\sigma_\phi$, $\sigma_{V^\prime}$, 
$\sigma_{V^{\prime\prime}}$, $\varphi_\rho$,
$\varphi_\phi$, and $\Delta M_\omega$. The latter parameter is a difference
between the fitted $\omega(782)$-mass value and the value obtained by 
SND~\cite{SNDom3pi} in the process $e^+e^-\to \pi^+\pi^-\pi^0$ 
on data of the 1998 and 2000 scans. The $\omega(782)$ width is fixed at 
the value from the same Ref.~\cite{SNDom3pi}. The mass and width for the 
$\phi(1020)$ are taken from the SND work~\cite{SNDphi3pi}, in
which data of the 1998 $\phi$-meson scan were used to study the processes 
$e^+e^-\to \pi^+\pi^-\pi^0$, $K^+K^-$, and $K_SK_L$. The $\rho(770)$
mass and width are fixed at the PDG values~\cite{pdg}.

The fit gives a small, consistent with zero, value of 
$\sigma_{V^{\prime\prime}}$. Therefore, in further analysis the 
model with $\sigma_{V^{\prime\prime}}=0$ is used. The fit 
describes data well, $\chi^2/ndf=40.9/55$, where $ndf$ is the number of
degrees of freedom. The following values of fitted parameters are obtained:
\begin{eqnarray}
&&\sigma_\rho=(0.485^{+0.55}_{-0.53}\pm0.025)\mbox{ nb},\,\, 
\sigma_\omega=(151.8\pm 1.3\pm 2.1)\mbox{ nb},\nonumber\\
&&\sigma_\phi=(5.53^{+1.00}_{-0.57}\pm 0.72)\mbox{ nb},\,\,
\sigma_{V^\prime}=(3.8^{+2.9}_{-2.4}\pm3.8)\mbox{ pb},\nonumber\\
&&\varphi_\rho=(-12.7\pm3.4\pm3.0)^\circ,\,\,
\varphi_\phi=(158^{+31}_{-18}\pm21)^\circ,\nonumber\\
&&\Delta M_\omega=53\pm42 \mbox{ keV}.\label{fit1}
\end{eqnarray}
The first error is statistical and the second is systematic. For the cross 
sections, the systematic uncertainty  includes the uncertainty of luminosity
determination (1.2\%), the uncertainty of the detection efficiency (0.6\%
for $\sigma_\rho$, $\sigma_\omega$, $\sigma_\phi$),
and uncertainties associated with the model of the $V^\prime$ 
energy-dependent width and inaccuracies of the resonance masses and widths.
It should be noted that a systematic uncertainty on $\sigma_\omega$   
is determined by the errors of the luminosity and detection efficiency, and
weakly depends on other sources. The main contribution to the 
$\sigma_\rho$ systematic uncertainty comes from inaccuracy of 
$\Gamma_\omega$. The uncertainties of the $V^\prime$ parameters dominate in
the $\sigma_\phi$, $\varphi_\rho$, and $\varphi_\phi$ systematic errors.
The fitted shift of the $\omega$ mass relative to the SND measurement
in the $e^+e^-\to \pi^+\pi^-\pi^0$ process~\cite{SNDom3pi}
is consistent with zero.
This parameter is allowed to float because its statistical uncertainty of 42
keV is lower than the systematic uncertainty of the $\omega$ mass (90 keV)
quoted in Ref.~\cite{SNDom3pi}.
To understand importance of the $V^\prime$ contribution, we perform
a fit with $\sigma_{V^\prime}=0$. The obtained $\chi^2$ value equal to
45.4 corresponds to a significance 
of 2.1 standard deviations for the $V^\prime$ contribution.  

\begin{figure}
\includegraphics[width=0.32\textwidth]{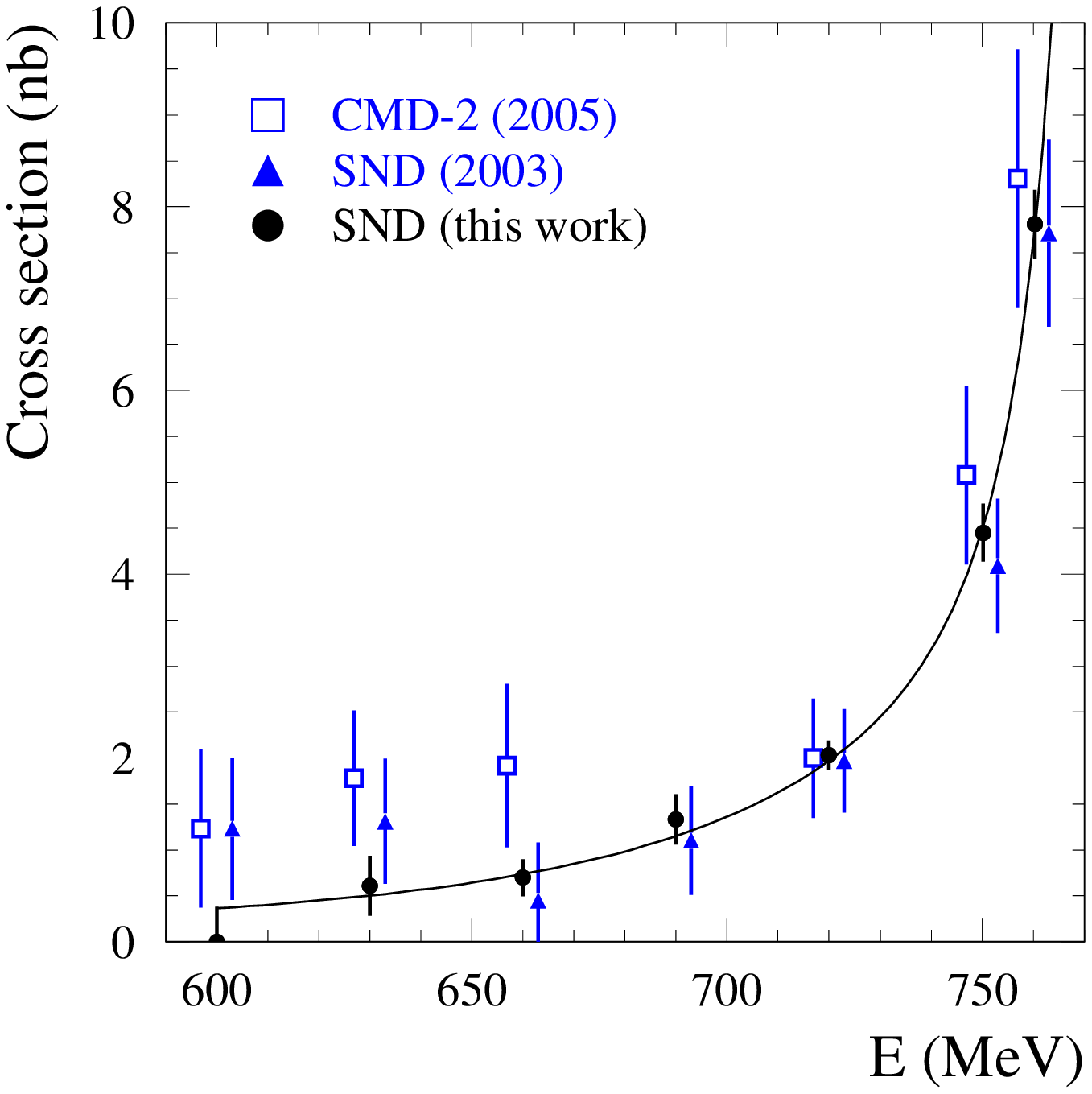}
\includegraphics[width=0.32\textwidth]{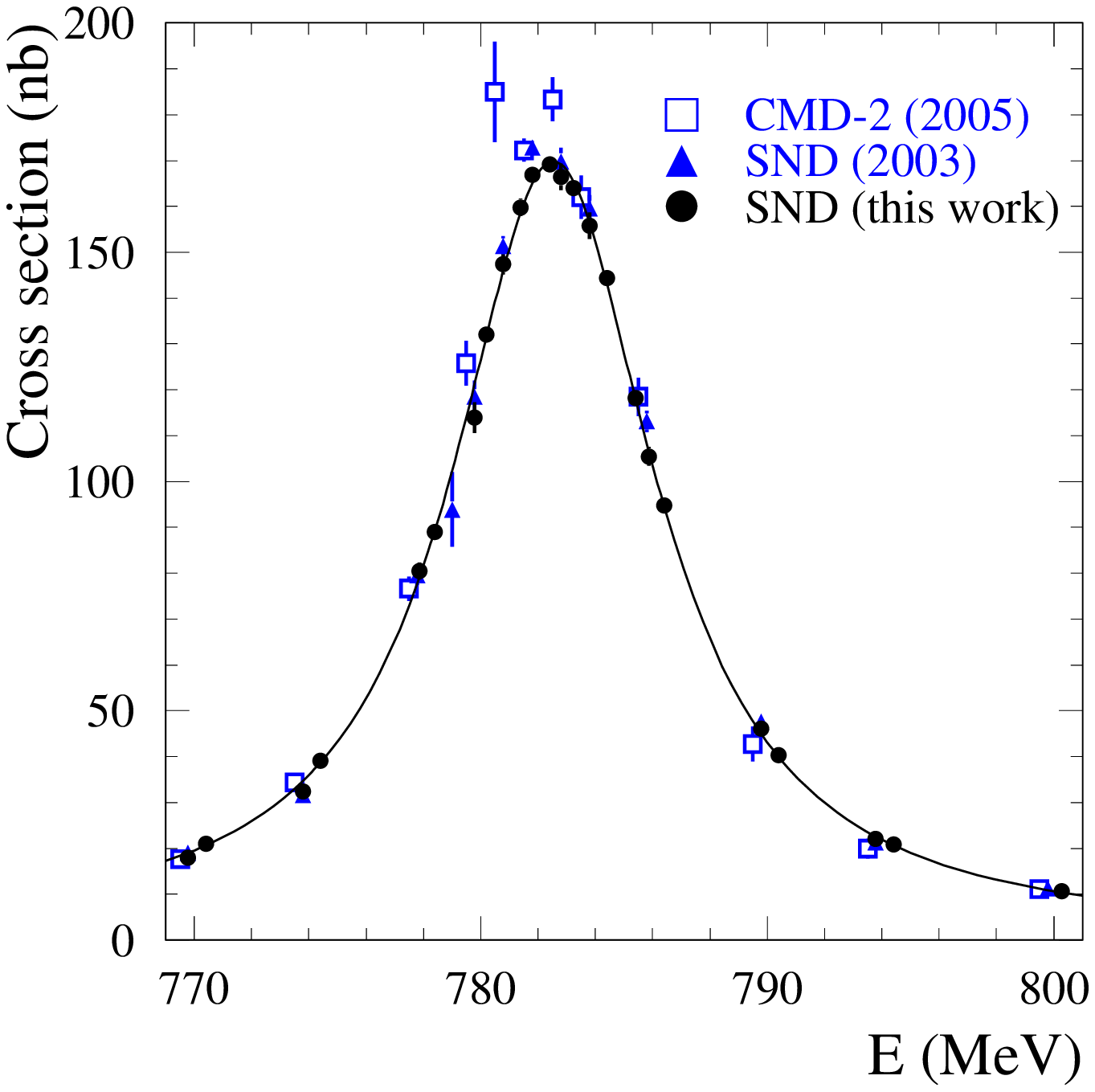}
\includegraphics[width=0.32\textwidth]{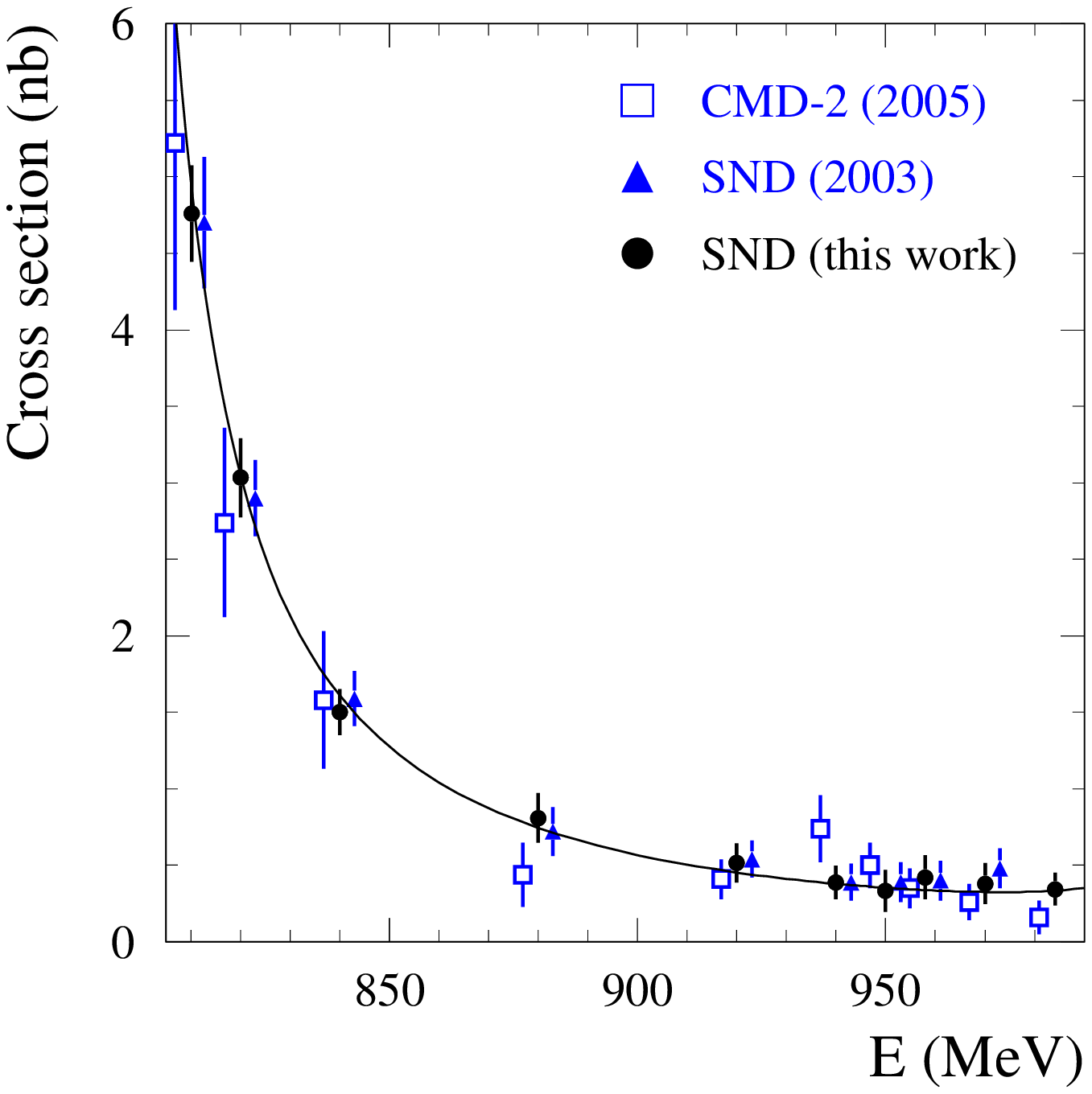}
\includegraphics[width=0.32\textwidth]{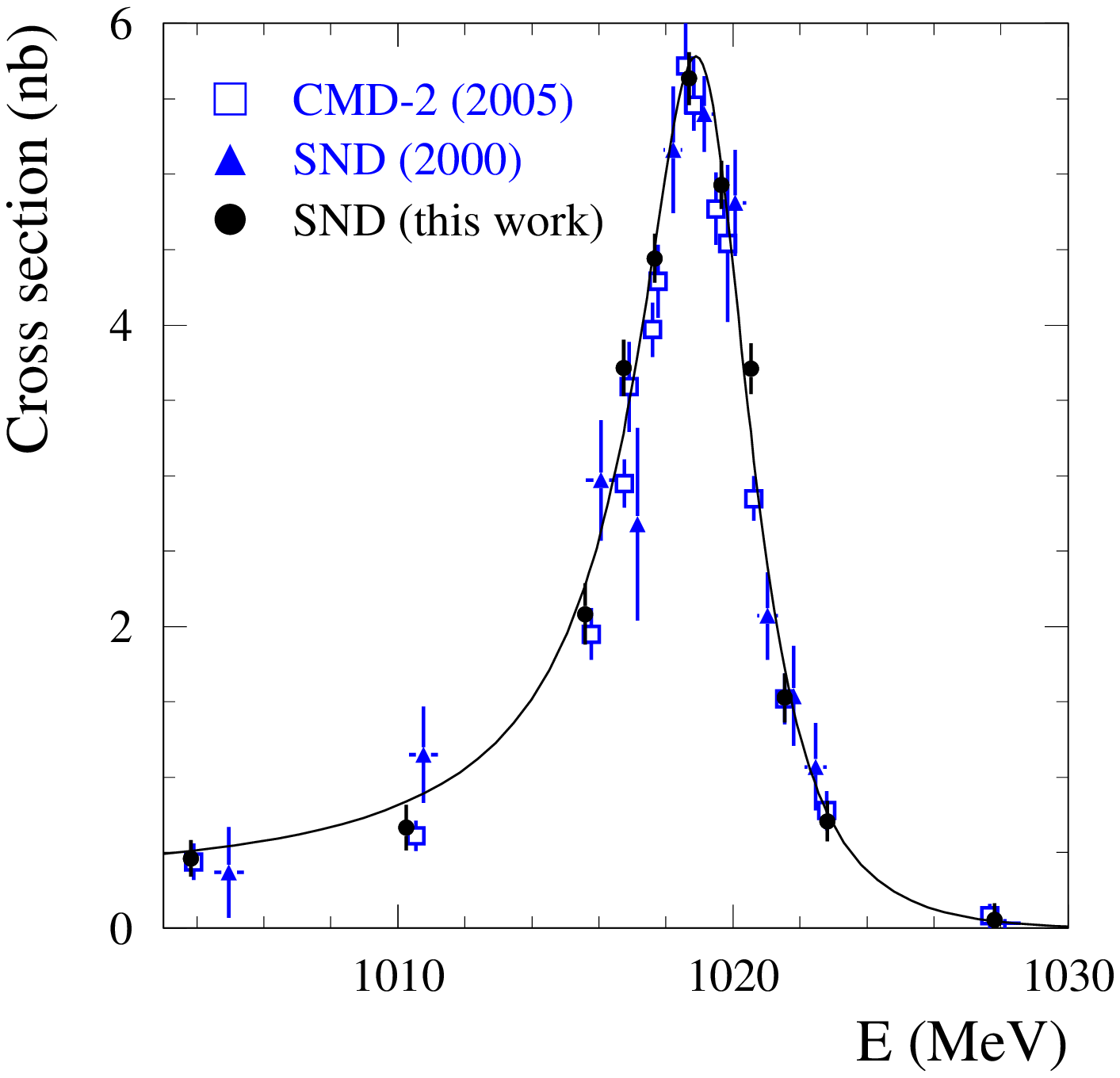}
\includegraphics[width=0.32\textwidth]{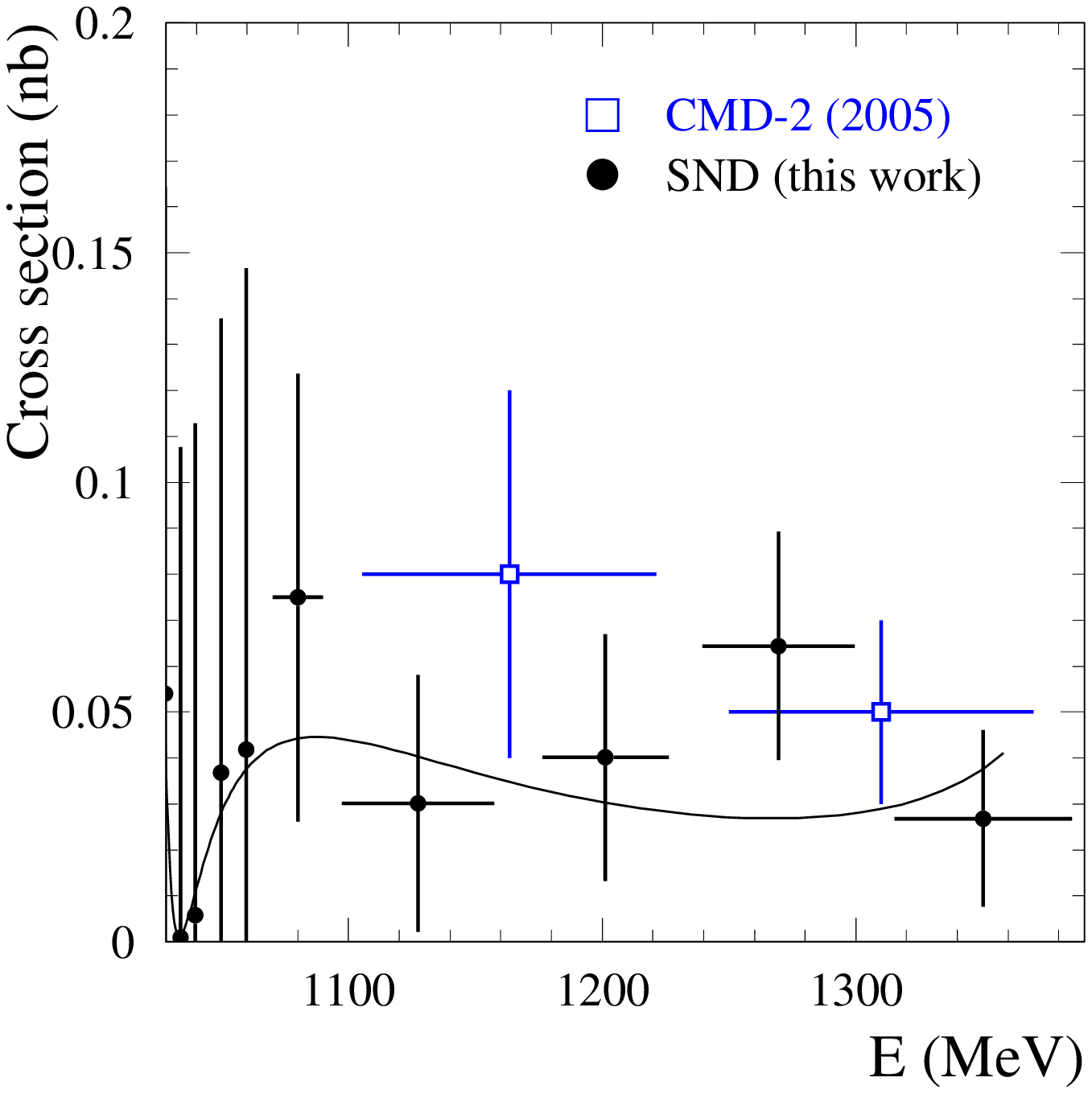}
\caption{The $e^+e^-\to \pi^0\gamma$ cross section measured in this work
in different energy regions in comparison with the previous most accurate
measurements: SND (2000)~\cite{snd1}, SND (2003)~\cite{snd2}, and
CMD-2 (2005)~\cite{cmd}. The curve is the result of the fit described 
in the text. In the energy region 1030--1100 GeV CMD-2 set upper limits
of 0.1--0.2 nb, which is not shown in the corresponding plot.
In the upper-left and upper-right plots the CMD-2 (2005) and SND (2003) data
points are shifted from their actual energies by $-2$ and $+2$ MeV,
respectively. Only statistical errors are 
shown. The systematic errors are 3.2\%, 3\%, and 6\% for 
SND (2000), SND (2003), and CMD-2 (2005) data, respectively. 
\label{fig5}}
\end{figure}

Substituting the fitted cross section $\sigma(s)$ into Eq.~(\ref{viscrs})
and Eq.~(\ref{viscrs2}), we calculate the radiative corrections
and the experimental values of the Born cross section. They are listed 
in Table~\ref{allres}. The radiative corrections are also calculated
with the different models of the $V^\prime$ and with the fit parameters 
varied within their uncertainties. The maximum deviation of a radiative 
correction from its nominal value is taken as an estimate of its uncertainty. 
The quoted errors on the Born cross section are statistical and systematic.
The latter includes the uncertainties in luminosity, detection
efficiency, and radiative correction, as well as the uncertainty 
associated with inaccuracy in energy setting. 
The Born cross section measured in this work is shown in Fig.~\ref{fig5} in 
different energy regions in comparison with the previous most accurate
measurements~\cite{snd1,snd2,cmd}.
The obtained cross section is in reasonable agreement with the previous
SND measurements~\cite{snd1,snd2}, but is more precise.
Above 0.76 GeV the CMD-2 and our data agrees within the 6\% systematic 
uncertainty~\cite{cmd} for most data points.
The exception is the point at 780.5 MeV with about 3$\sigma$ deviation
from the fitting curve. Below 0.76 GeV there is a systematic difference 
between our and CMD-2 measurements.

\section{Discussion\label{finres}}
From the measured peak cross sections [Eq.~(\ref{fit1})] we calculate the
products of branching fractions for the $\rho$, $\omega$, and $\phi$
mesons:
\begin{eqnarray}
B(\rho\to\pi^0\gamma)B(\rho\to e^+e^-)&=&
(1.98\pm0.22\pm0.10)\times10^{-8},\nonumber\\
B(\omega\to\pi^0\gamma)B(\omega\to e^+e^-)&=&
(6.336\pm0.056\pm0.089)\times10^{-6},\nonumber\\
B(\phi\to\pi^0\gamma)B(\phi\to e^+e^-)&=&
(3.92_{-0.40}^{+0.71}\pm0.51)\times 10^{-7}.\label{prod1}
\end{eqnarray}
Our results agree with previous measurements of these 
parameters. The accuracies of the products for the $\rho$ and $\omega$
mesons are improved by a factor of about 2 compared with the 
most precise previous measurement by SND~\cite{snd2}.
For the $\phi$ meson, our total uncertainty is about 20\%. This is due 
to a strong correlation between  $\sigma_\phi$ and $\varphi_\phi$.
Our uncertainty is significantly, by a factor of 2, larger than 
the uncertainty of previous measurements~\cite{snd1,cmd}.
Since the accuracy of the $e^+e^- \to\pi^0\gamma$ cross section measured 
in this work is better than that
of the previous measurements, we conclude that the systematic uncertainty
on $B(\phi\to\pi^0\gamma)B(\phi\to e^+e^-)$ was previously underestimated.

The phase $\varphi_\rho$ can be calculated from  
$B(\omega\to\pi^+\pi^-)$~\cite{pdg} assuming that this decay is fully 
determined by electromagnetic 
$\rho-\omega$ mixing~\cite{mix1,mix2}. It is found to be 
$(- 13.5\pm 0.6)^\circ$ and agrees well with our measurement
$\varphi_\rho=(-12.8\pm3.5\pm3.0)^\circ$. It is expected that the phase 
$\varphi_\phi$ is close to the same phase measured in the
$e^+e^-\to \pi^+\pi^-\pi^0$ reaction
$\varphi_\phi^{3\pi}=(163\pm7)^\circ$~\cite{SNDom3pi}.
Since our result on $\varphi_\phi$ does not contradict this expectation, we 
can improve the accuracy of the $\phi$-meson peak cross section by fixing
the parameter $\varphi_\phi$ at the value obtained from
$e^+e^-\to \pi^+\pi^-\pi^0$. The fit yields the following value for
the product of the branching fractions:
\begin{eqnarray}
B(\phi\to\pi^0\gamma)B(\phi\to e^+e^-)&=&
(4.04\pm0.09\pm0.19)\times10^{-7},\label{prod2}
\end{eqnarray}
where the systematic error is dominated by the uncertainty on
$\varphi_\phi^{3\pi}$.

Using the measured product $B(\omega\to\pi^0\gamma)B(\omega\to e^+e^-)$
and the PDG value $B(\omega\to\pi^+\pi^-\pi^0)B(\omega\to e^+e^-)=
(6.38\pm0.10)\times10^{-5}$~\cite{pdg}, we calculate the ratio
\begin{eqnarray}
\frac{B(\omega\to\pi^0\gamma)}{B(\omega\to\pi^+\pi^-\pi^0)}&=&
0.0992\pm0.0023,
\label{brat}
\end{eqnarray}
which disagrees (by $3.4\sigma$) with the KLOE measurement of
the same parameter $0.0897\pm0.0016$~\cite{kloe}. The KLOE Collaboration
obtained the ratio of the $\omega$ branching fractions from the ratio
of the cross sections for $e^+e^-\to \omega\pi^0\to \pi^0\pi^0\gamma$ and
$e^+e^-\to \omega\pi^0\to \pi^+\pi^-\pi^0\pi^0$ measured near
the $\phi(1020)$ resonance. This technique was suggested by the SND 
Collaboration~\cite{SNDrat}. The SND result 
$B(\omega\to\pi^0\gamma)/B(\omega\to\pi^+\pi^-\pi^0)=0.0994\pm0.0052$
agrees well with Eq.~(\ref{brat}) and differs from the  KLOE measurement
by 1.8$\sigma$.

It is instructive to calculate the product 
$B(\omega\to\pi^+\pi^-\pi^0)B(\omega\to e^+e^-)$ using the
KLOE value of $B(\omega\to\pi^0\gamma)/B(\omega\to\pi^+\pi^-\pi^0)$ 
and our result on $B(\omega\to\pi^0\gamma)B(\omega\to e^+e^-)$.
The result 
\begin{equation}
B(\omega\to\pi^+\pi^-\pi^0)B(\omega\to e^+e^-)_{\rm KLOE+this~work}=
(7.06\pm0.17)\times10^{-5}
\label{prodkloe}
\end{equation}
exceeds the PDG value by 3.4$\sigma$. It should be noted that the PDG value 
is the average of eight measurements, which are in reasonable agreement with 
each other. 

The KLOE measurement strongly influences current PDG values of $\omega$ meson 
parameters. Therefore, we calculate $\omega$ meson parameters 
based on our measurement $B(\omega\to\pi^0\gamma)B(\omega\to e^+e^-)$,
the PDG values of $B(\omega\to\pi^+\pi^-\pi^0)B(\omega\to e^+e^-)$,
and branching fractions of other decays, which sum is equal to
$0.0165\pm0.0013$. The following parameters are 
obtained:
\begin{eqnarray}
B(\omega\to\pi^0\gamma)&=&(8.88\pm0.18)\%, \nonumber\\
B(\omega\to\pi^+\pi^-\pi^0)&=&(89.47\pm0.18)\%, \nonumber\\
B(\omega\to e^+e^-)&=&(7.13\pm0.10)\times10^{-5},
\end{eqnarray}
which can be compared with the corresponding PDG values $(8.28\pm0.28)\%$, 
$(89.2\pm0.7)\%$, $(7.28\pm0.14)\times10^{-5}$.
As expected, our result for $B(\omega\to\pi^0\gamma)$ strongly
differs from the PDG value.

Using the PDG values for $B(\rho\to e^+e^-)$ and $B(\phi\to e^+e^-)$ we
calculate the branching fractions
\begin{eqnarray}
B(\rho\to\pi^0\gamma)&=&(4.20\pm0.47\pm0.22)\times10^{-4},\nonumber\\
B(\phi\to\pi^0\gamma)&=&(1.367\pm0.030\pm0.065)\times10^{-3}
\end{eqnarray}
Our result on $B(\phi\to\pi^0\gamma)$ agrees with the PDG
value $(1.27\pm0.06)\times10^{-3}$ and has comparable accuracy. For
$B(\rho\to\pi^0\gamma)$ our result is lower than the PDG value
$(6.0\pm0.8)\times10^{-4}$ by $1.8\sigma$, but agrees with the branching
fraction for the charged
$\rho$ decay $B(\rho^\pm\to\pi^\pm\gamma)=(4.5\pm0.5)\times10^{-4}$.

\section{Summary}
The cross section for the process
$e^+e^-\to\pi^0\gamma$  has been
measured in the energy range of 0.60--1.38 GeV with the SND detector 
at the VEPP-2M $e^+e^-$ collider. This is the most accurate
measurement of the cross section.
Data on the cross section are well fitted with
the VMD model with the $\rho(770)$, $\omega(782)$, $\phi(1020)$, and
an additional resonance describing a total contribution of
the $\rho(1450)$ and $\omega(1420)$ resonances. From this fit we
have determined the products of branching fractions
\begin{eqnarray}
B(\rho\to\pi^0\gamma)B(\rho\to e^+e^-)&=&
(1.98\pm0.22\pm0.10)\times10^{-8},\nonumber\\
B(\omega\to\pi^0\gamma)B(\omega\to e^+e^-)&=&
(6.336\pm0.056\pm0.089)\times10^{-6},\nonumber\\
B(\phi\to\pi^0\gamma)B(\phi\to e^+e^-)&=&
(4.04\pm0.09\pm0.19)\times10^{-7},
\end{eqnarray}
and the branching fractions
\begin{eqnarray}
B(\rho\to\pi^0\gamma)&=&(4.20\pm0.52)\times10^{-4},\nonumber\\
B(\omega\to\pi^0\gamma)&=&(8.88\pm0.18)\%, \nonumber\\
B(\phi\to\pi^0\gamma)&=&(1.367\pm0.072)\times10^{-3}.
\end{eqnarray}
Our measurements for the $\rho\to\pi^0\gamma$ and $\omega\to\pi^0\gamma$ 
branching fractions have accuracies better than those for the
PDG values~\cite{pdg}. Our result for $B(\rho\to\pi^0\gamma)$ is lower 
than the PDG value by $1.8\sigma$, but agrees with the branching
fraction for the charged $\rho$ decay. For the $\omega$, the values of the 
three directly measured parameters,
$B(\omega\to\pi^0\gamma)B(\omega\to e^+e^-)$, 
$B(\omega\to\pi^+\pi^-\pi^0)B(\omega\to e^+e^-)$, and
$B(\omega\to\pi^0\gamma)/B(\omega\to\pi^+\pi^-\pi^0)$, contradict
each other. With our measurement, the level of disagreement between them
reaches 3.4$\sigma$.
The result for the $\phi\to\pi^0\gamma$ has an accuracy comparable
with that of the PDG value~\cite{pdg}. It has been obtained
assuming that the relative phase between the $\phi$ and $\omega$ meson
amplitudes is equal to the phase determined in the $e^+e^-\to \pi+\pi^-\pi^0$  
process, $\varphi_\rho^{3\pi}=(163\pm7)^\circ$~\cite{SNDom3pi}. Without this 
assumption, $B(\phi\to\pi^0\gamma)$ is determined with about 20\%
uncertainty.

The results presented in this paper supersede our previous 
measurement~\cite{snd2} based on a part of data collected by SND at 
VEPP-2M below 1 GeV. 

We thank S.I. Eidelman for fruitful discussions.
This work is supported by the Ministry of Education and Science of the
Russian Federation and the RFBR grant 
No. 13-02-00375-a.

 \end{document}